# Advancing Memristive Analog Neuromorphic Networks: Increasing Complexity, and Coping with Imperfect Hardware Components

F. Merrikh Bayat[1], M. Prezioso[1], B. Chakrabarti[1], I. Kataeva[2], and D. B. Strukov[1]
[1]UCSB, Santa Barbara, CA 93106-9560, U.S.A.
[2]Research Laboratories, DENSO CORP., 500-1 Minamiyama, Komenoki-cho, Nisshin, Japan 470-0111

***Abstract*** - We experimentally demonstrate classification of 4×4 binary images into 4 classes, using a 3-layer mixed-signal neuromorphic network ("MLP perceptron"), based on two passive 20×20 memristive crossbar arrays, board-integrated with discrete CMOS components. The network features 10 hidden-layer and 4 output-layer analog CMOS neurons and 428 metal-oxide memristors, i.e. is almost an order of magnitude more complex than any previously reported functional memristor circuit. Moreover, the inference operation of this classifier is performed entirely in the integrated hardware. To deal with larger crossbar arrays, we have developed a semi-automatic approach to their forming and testing, and compared several memristor training schemes for coping with imperfect behavior of these devices, as well as with variability of analog CMOS neurons. The effectiveness of the proposed schemes for defect and variation tolerance was verified experimentally using the implemented network and, additionally, by modeling the operation of a larger network, with 300 hidden-layer neurons, on the MNIST benchmark. Finally, we propose a simple modification of the implemented memristor-based vector-by-matrix multiplier to allow its operation in a wider temperature range.

## I. Introduction

Several types of emerging nonvolatile memory devices are now being actively investigated for their use in fast and energy-efficient analog and mixed-signal neuromorphic networks [1-8]. For relatively immature technologies, such as RRAM devices [9] (also called memristors [10]), and even more mature PCM cells [6], the best previously reported results were obtained, to the best of our knowledge, combining experimental devices with external computers (for example, by reading from and writing to one device at a time [6]) to emulate the functionality of the whole system.

One key result of our work is an experimental demonstration of a fully functional, board-integrated, mixed-signal memristor-based neural network of a complexity much higher than those reported previously. Another important result is an experimental verification of several in-situ and ex-situ approaches to defect- and variation-tolerant memristor training. Specifically, we have analyzed training of a pattern classifier based on a firing-rate neural network (MLP perceptron), which is very efficient for high-performance implementation of deep-learning algorithms [11].

Our focus is on using passive metal-oxide memristor crossbar arrays, with crosspoint devices of a very small chip footprint (determined only by the overlap area of the wire electrodes). Such memristors may be scaled down below 10 nm without sacrificing their endurance, retention, and tuning accuracy, with some of the properties (e.g., the ON/OFF conductance ratio) being actually improved [12]. Moreover, these devices are naturally suitable for 3D integration [4, 13], which may be instrumental for keeping all the data required for deep learning locally, and thus cutting dramatically the energy and latency overheads of off-chip communications.

## II. Memrsitive Crossbar Arrays

20×20 crossbar arrays, with 200-nm lines separated by 400-nm gaps (Fig. 1), and a $Pt/Al_2O_3/TiO_{2-x}/Ti/Pt$ memristor at each crosspoint, were fabricated using a technique similar to that reported in Refs. 2, 3. To speed up the memristor forming procedure, a setup for its automation was developed (Fig. 2). The setup was used for early screening of defective samples, and has allowed a successful forming and testing of numerous crossbar arrays. As Fig. 3 shows, we have explored forming using both current and voltage stress pulses, but have not detected much difference between the two, likely due to larger parasitics of these larger arrays.

Similarly to our previously reported results [2, 3], the crossbar devices have relatively good uniformity, with a spread of the set and reset voltages narrow enough (Fig. 3) to allow a precise adjustment of each memristor of the whole array (Fig. 4). To our knowledge, this is the first report of such a precise adjustment on this integration scale; for example, Ref. 14 reported a less precise tuning, performed for smaller (8×8 device) fragments of a larger array, with all the remaining devices always kept in the high-resistive state.

## III. Multilayer Perceptron Implementation

The implemented MLP perceptron is fed with 16 binary inputs encoding 4×4 B/W pixels with ±0.2 voltages, and consists of 10 hidden and 4 output layer neurons, connected with two 20×20-memristor arrays (Figs. 5a,b). With additional bias inputs, 17×20 and 11×8 portions of the arrays were used to implement differential synaptic weights, with memristor conductances $G^{\pm}$ in the range [10 μs, 100 μs] in the first and second crossbars. The neurons, as well as the circuitry for weight adjustment, were implemented with discrete CMOS components. All components of the system, were integrated on a printed circuit board (Figs. 5c, d).

Figure 5e shows the design of a hidden-layer neuron, which consists of two opamps, computing a pair of differential voltages $R_F\Sigma_i G_i^{\pm} V_i$, where $R_F$ = 2 kΩ is feedback resistance, by enforcing the virtual ground condition on the incoming crossbar lines. This pair of voltages is then fed into a third opamp, which computes the difference between the inputs, and clips the output voltage, keeping it between the voltage supply rails, thus effectively implementing a piece-linear activation function with a low-voltage slope of 10 and saturation at ±5 V. This output is scaled down, with one more opamp circuit, to be within at most ±0.2 V, to avoid disturbing the state of memristors in the second layer.

## IV. Coping with Imperfect Hardware

The perceptron was trained to classify stylized letter patterns (Fig. 6), using four alternative approaches (Fig. 7). In the simplest ex-situ training (Figs. 7a, 8), the weights are computed in a "precursor" external computer, assuming perfect on-chip hardware, and then are "imported" into the crossbars. This method has the lowest on-chip hardware overhead and is suitable for the most popular applications of neuromorphic networks. However, though the write-verify algorithm circumvents the problem of threshold variations in memristors, such ex-situ approach cannot cope with other imperfections in the hardware, such as stuck-on or stuck-off defects (Fig. 9b) and the device *I-V* curve asymmetry (Figs. 1d, 9a). The ex-situ classification fidelity may be significantly improved, e.g., from 95% (Fig. 8a) to 100% (Fig. 9c) on the 4-class pattern set (Fig. 6), by detecting various defects and then using this information at the precursor training (Fig. 7b). A potential drawback of such defect-aware ex-situ scheme is that the chip-specific precursor training may not be suitable for some applications - e.g., when training takes too much time.

An apparent alternative is the in-situ training, performed directly in a hardware (Figs. 7c, 10). With a supporting on-chip training circuitry, the in-situ approach might be utilized to implement on-line learning, making it suitable for a broader range of applications. (In our demonstration, similar to that described in Refs. [2, 3], some stages of in-situ training were assisted by an external computer.) However, in our experiments this approach, in its batch-mode, fixed-amplitude version [2], has provided an inferior fidelity of ~70% for a 3-class pattern set, compared to the 100% fidelity for both ex-situ approaches for the same task. The main reason, as confirmed by simulations based on a simple dynamic model [3] (Fig. 10b, c), are large variations of the switching thresholds, which are more effectively coped by the tuning algorithm at the ex-situ training. The in-situ training fidelity could potentially be improved using a variable-amplitude mapping scheme [2], which was not possible in the current design due to fixed voltage inputs. A hybrid approach (Fig. 7d) - first initializing weights to the values prescribed by ex-situ training, and then readjusting them with in-situ training, leads to a much better classification fidelity. Actually, experimental results show that with more artificially injected defects, the fidelity may be much higher than that at the purely ex-situ approach (Figs. 11a). This fact is also confirmed via simulations of a much larger network (Figs 12a, b). The simulation results shown in Figs. 12c-e confirm that other defect types are manageable in larger networks.

Finally, practical memristive hardware should be able to operate correctly under wide temperature ranges. For the considered circuits (Fig. 5b), the change in memristor conductance can be compensated by utilizing memristor as a feedback and mapping to a higher conductances (Fig. 13).

## Acknowledgment

This work was supported by AFOSR under MURI grant FA9550-12-1-0038, by DARPA under contract HR0011-13-C-0051UPSIDE via BAE Systems, Inc., by NSF grant CCF-1528205, and by the DENSO CORP., Japan. Useful discussions with P.-A. Auroux, J. Edwards, and K. K. Likharev are gratefully appreciated.

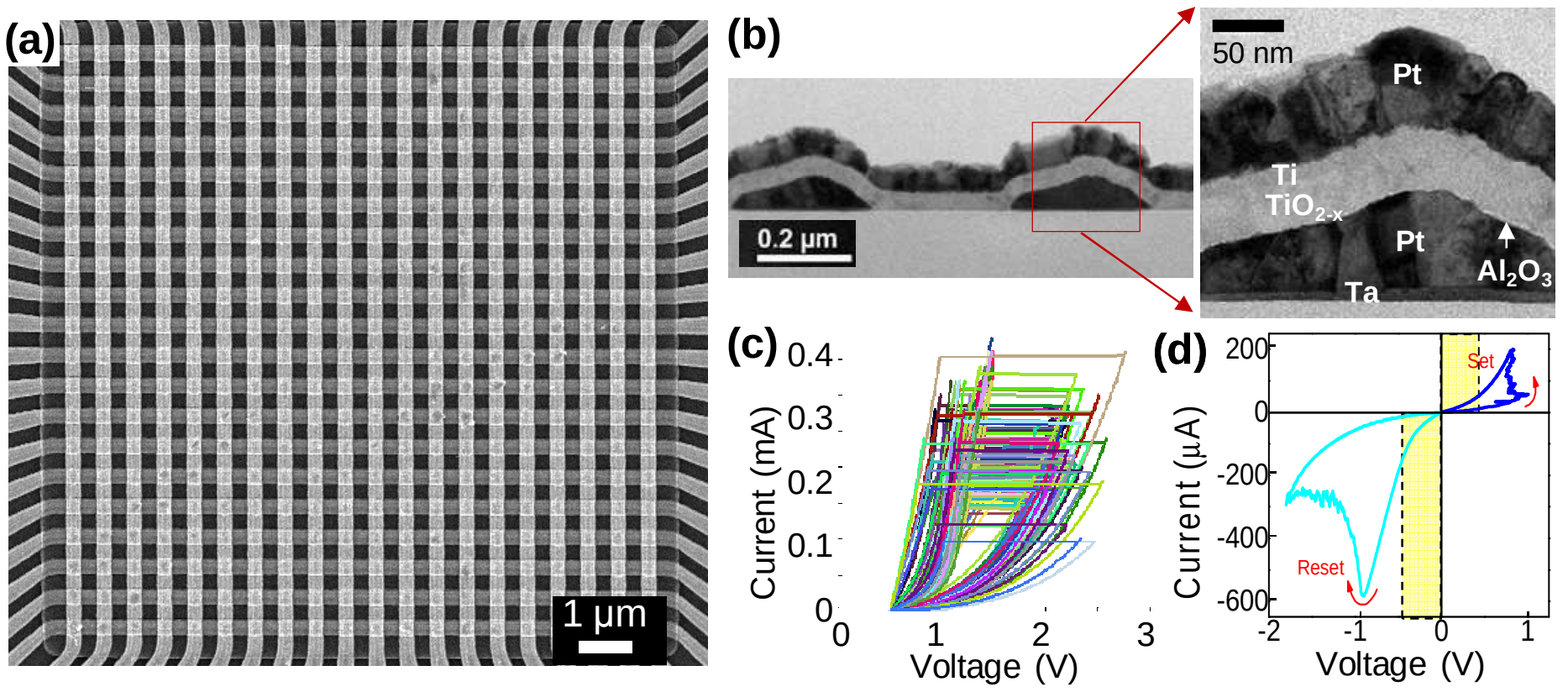

Fig. 1. 20×20 crossbar circuit with integrated Pt/$Al_2O_3$/$TiO_{2-x}$/Ti/Pt memristors: (a) a top-view SEM and (b) cross-section TEM images. (c) All forming *I-V* curves for one crossbar, and (d) a typical switching *I-V* curve, with its asymmetry clearly visible.

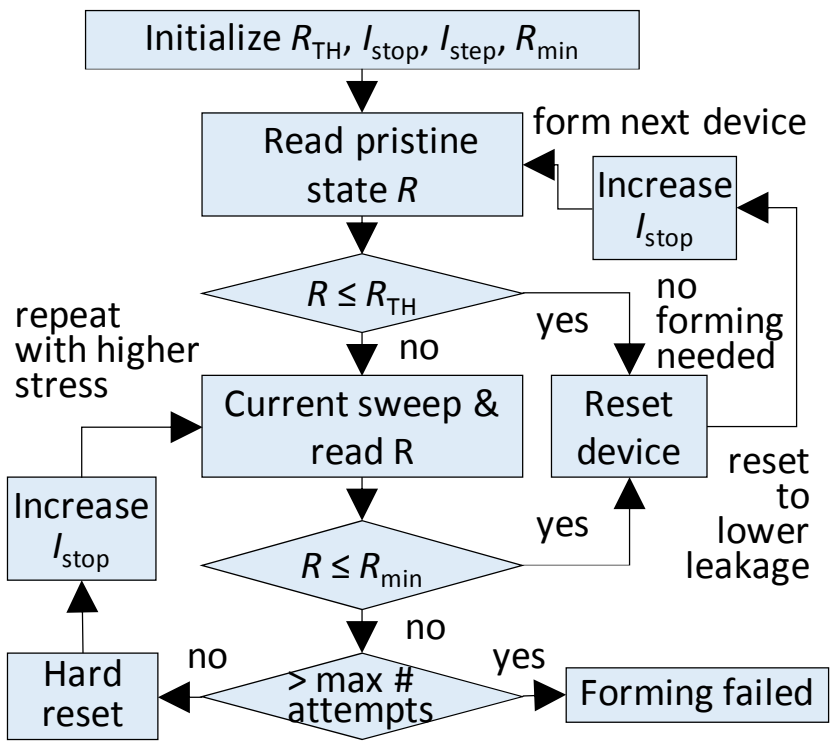

Fig. 2. Flow diagram of the automatic memristor forming procedure. The value of $I_{stop}$ was so far adjusted manually after the failure to form a device automatically (in ~10% of all cases).

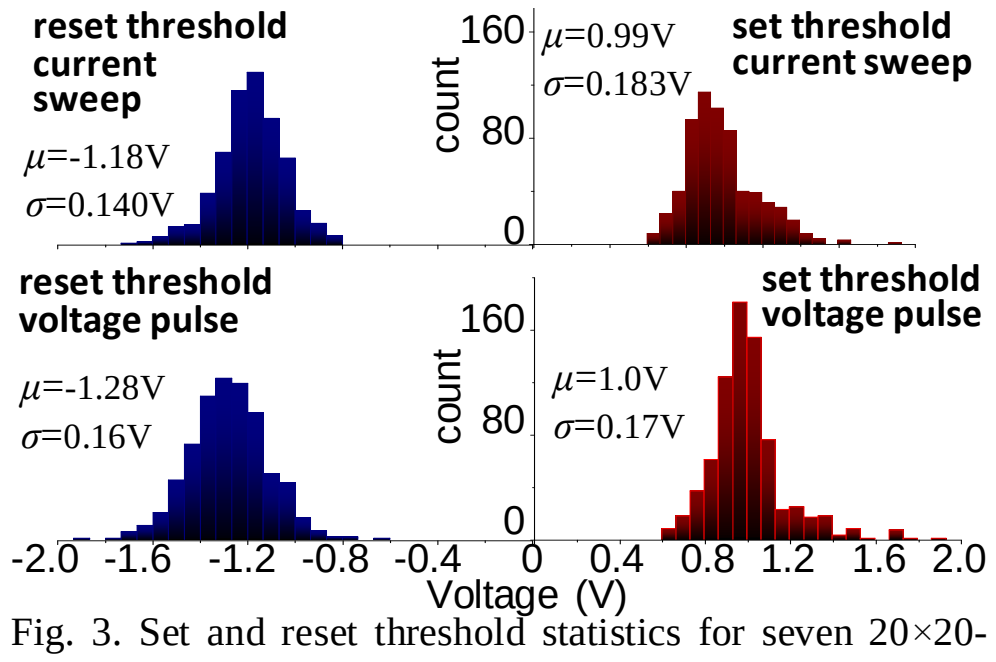

Fig. 3. Set and reset threshold statistics for seven 20×20-device arrays at memristor switching with current and voltage pulses. The set/reset thresholds are defined as the smallest voltages at which the device resistance is increased/decreased by more than 5% at the application of a voltage or current pulse of the corresponding polarity.

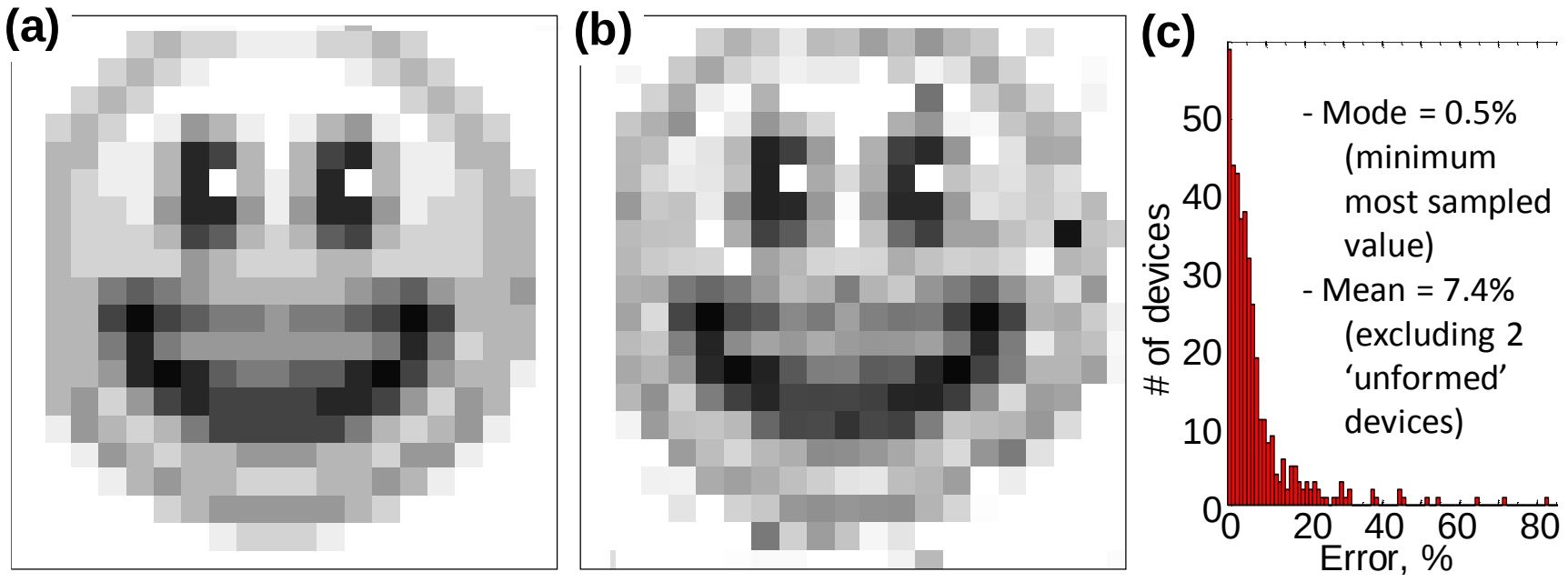

Fig. 4. High precision tuning in 20×20 memristive crossbar: (a) the desired "smiley face" pattern, quantized to 256 gray levels. (b) The actual resistance values measured after tuning all devices with the nominal 5% accuracy, using the automated tuning algorithm, and (c) the corresponding statistics of the tuning errors. On panel (b), the white / black pixels correspond to 84 kΩ / 7kΩ at 0.2 V bias.

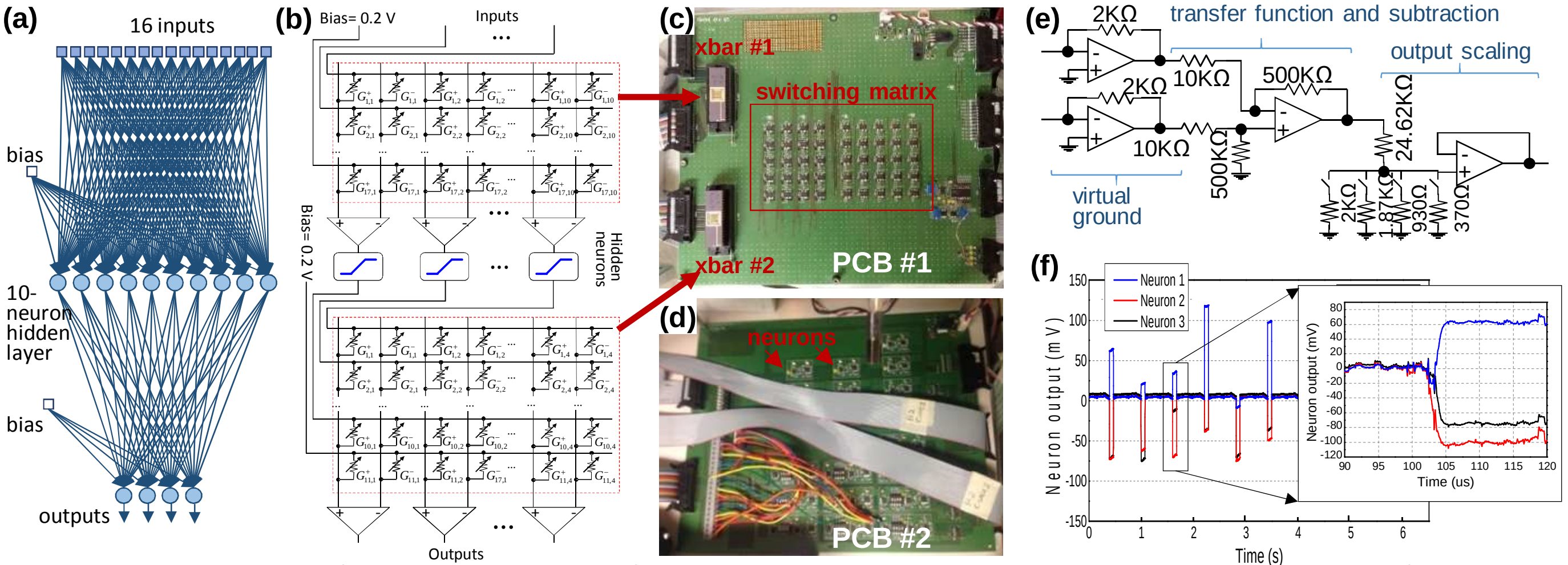

Fig. 5. Multilayer perceptron classifier: (a) Graph representation of the implemented network and (b) its equivalent electrical circuit. (c, d) Photos of the two printed circuit boards hosting (PCB #1) wire-bonded memristive crossbar chips and the memristor tuning circuitry, and (PCB #2) discrete CMOS neurons. (e) Equivalent circuit of a hidden layer neuron based on discrete CMOS opamps. The output layer neurons are implemented without the output scaling (the last opamps) and the 10 KΩ pulldown resistor in the second-stage opamp. (f) Typical output signal dynamics during classification; note the few-microsecond signal time of the operation.

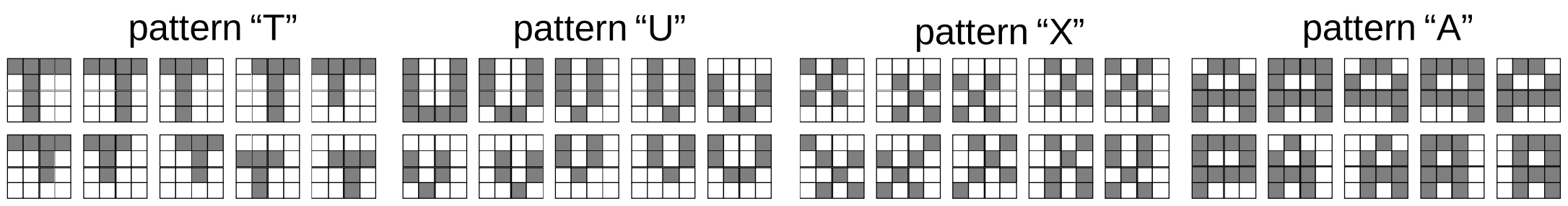

Fig. 6. 40 patterns, of 4 classes, used for the classifier training. 640 test patterns were formed by flipping one pixel in each training pattern.

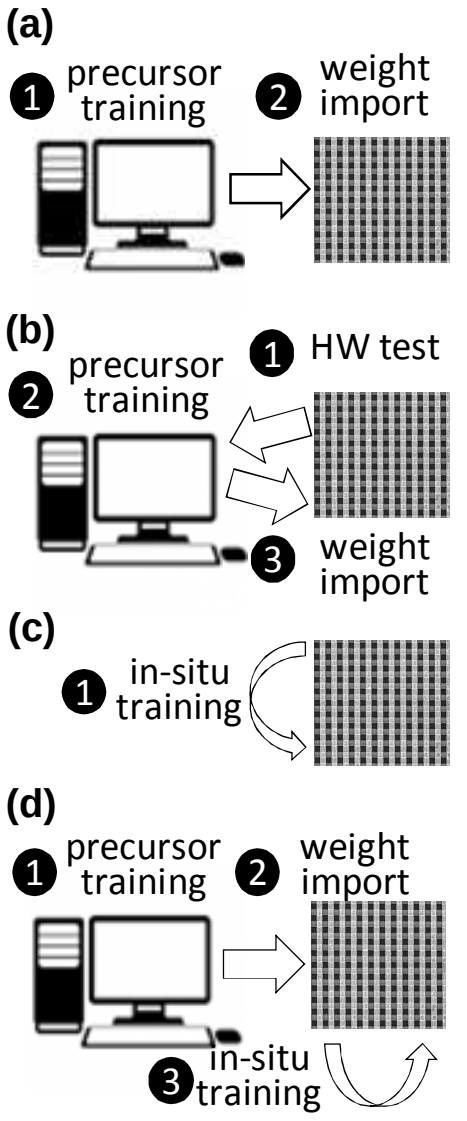

| | PROS | CONS |
|---|---|---|
| (a) | - Lowest HW overhead<br>- Good fidelity | - Poor imperfection tolerance<br>- Off-line learning |
| (b) | - Best imperfection tolerance<br>- Low HW overhead<br>- Best fidelity | - Poorly scalable step 1 (HW test)<br>- Off-line learning<br>- Chip-specific training |
| (c) | - Best imperfection tolerance<br>- Suitable for on-line learning | - High HW overhead<br>- Suboptimal fidelity<br>- Long training times<br>- Chip-specific training |
| (d) | - Best imperfection tolerance<br>- Suitable for on-line learning | - High HW overhead |

Fig. 7. Training approaches to cope with impecfect hardware: (a) ex-situ, (b) defect-aware ex-situ, (c) in-situ, and (d) hybrid.

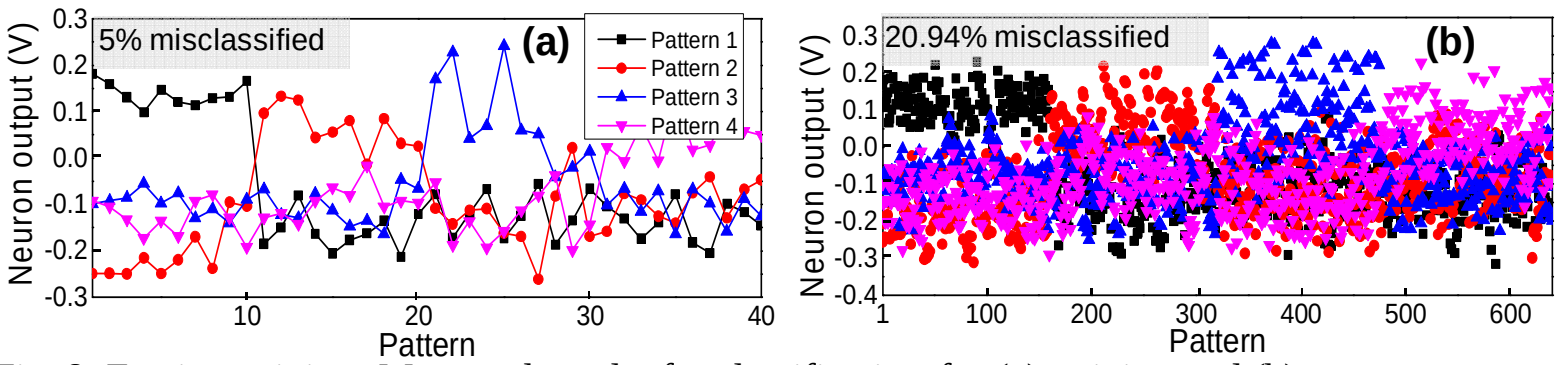

Fig. 8. Ex-situ training: Measured results for classification for (a) training and (b) test patterns.

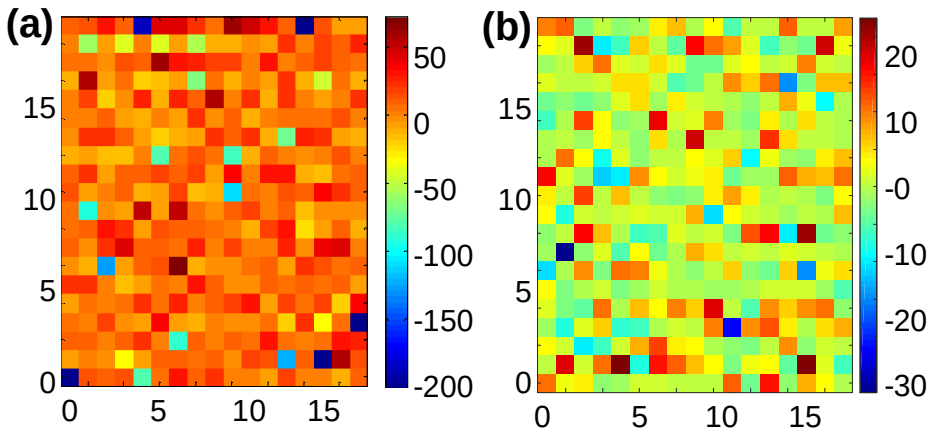

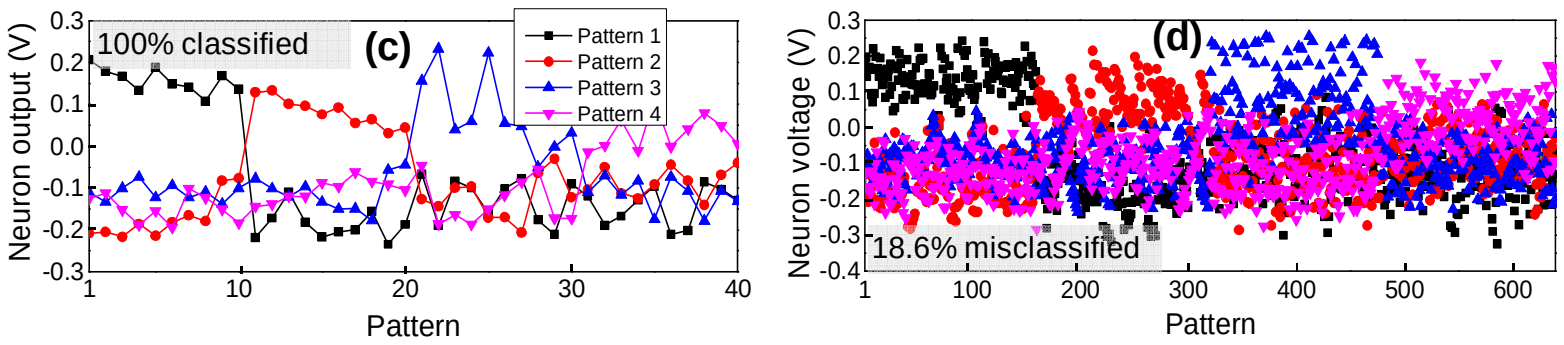

Fig. 9. Defect-aware ex-situ training: (a) the tuning error map and (b) the current (@ ±0.2 V) asymmetry map for the used 17×20 portion of the crossbar. Both panels show color-coded percentages. (c, d) Classification results for (c) training and (d) test patterns.

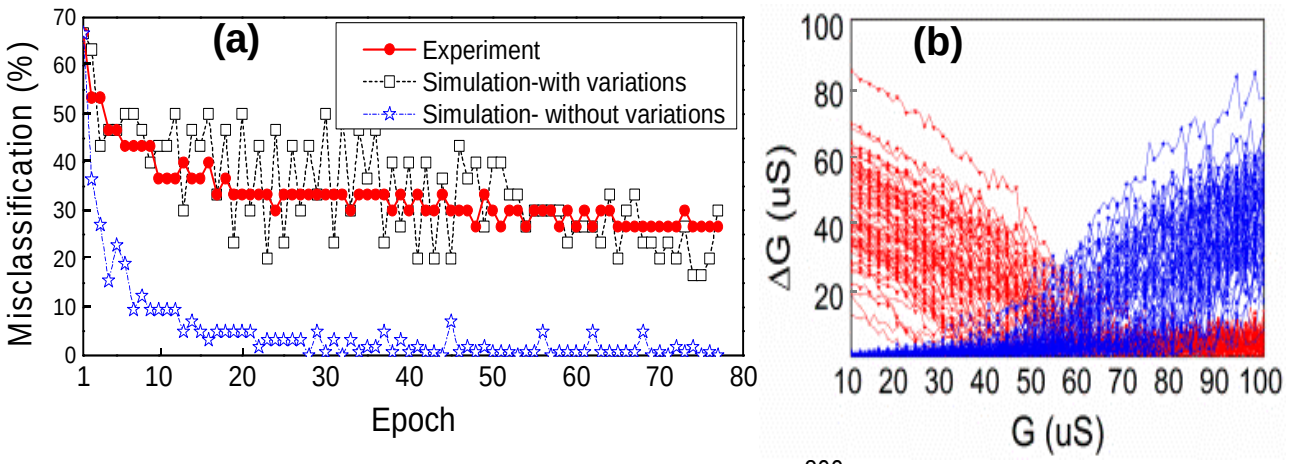

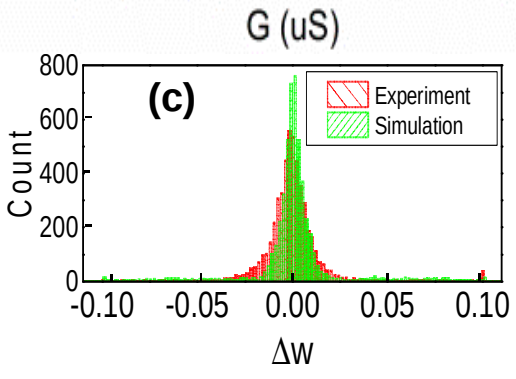

Fig. 10. In-situ training for 3-pattern classification: (a) Measured and simulated error decay dynamics during training. (b) Example of switching kinetics obtained from the model. (c) Comparison of the weight adjustment distributions computed for all weights over the entire batch training.

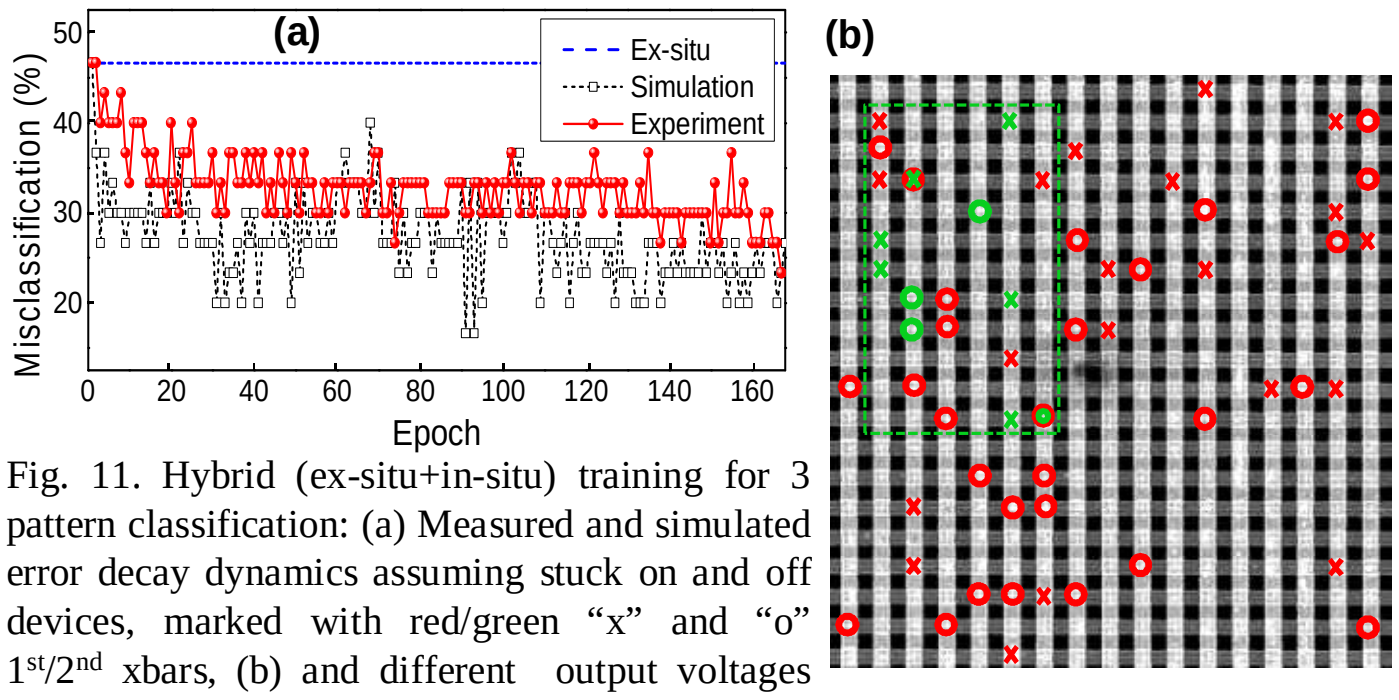

Fig. 11. Hybrid (ex-situ+in-situ) training for 3 pattern classification: (a) Measured and simulated error decay dynamics assuming stuck on and off devices, marked with red/green "x" and "o" 1st/2nd xbars, (b) and different output voltages (±0.4 V in 1st, 7th, 8th, ±0.1 V in 2nd, 4th, 5th, and ±0.2 V in the rest) in hidden-layer neurons.

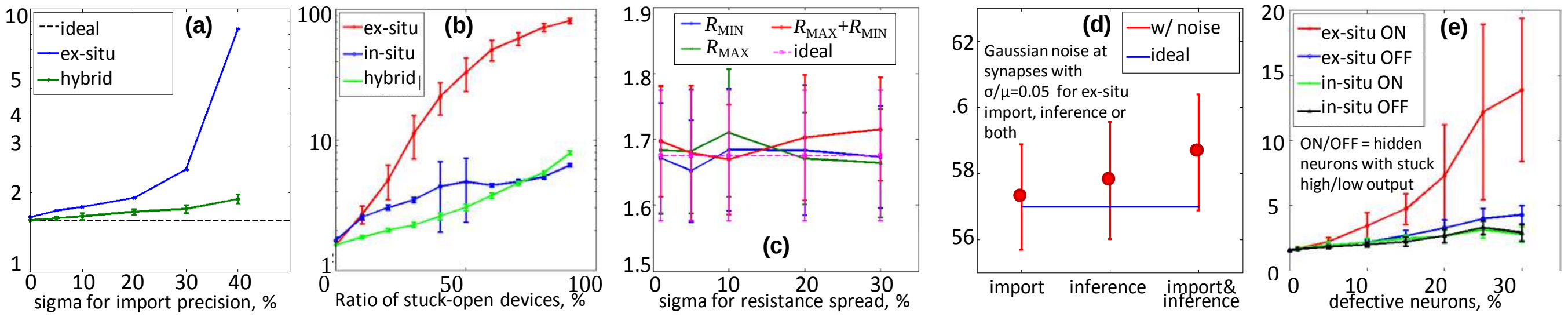

Fig. 12. Modeling of imperfect hardware effects at different training approaches. Classification rate degradation with respect to (a) finite weight import accuracy, (b) stuck-on and stuck-off memristors, (c) variations in maximum $R_{OFF}$ and $R_{ON}$ during in-situ training, (d) synaptic noise during import, inference, or both, and (e) stuck-on-high and stuck-on-low neurons. On all panels, the vertical axis is the classification error rate in percent; all data are for a 3-layer perceptron with 300 hidden layer neurons, and MNIST benchmark.

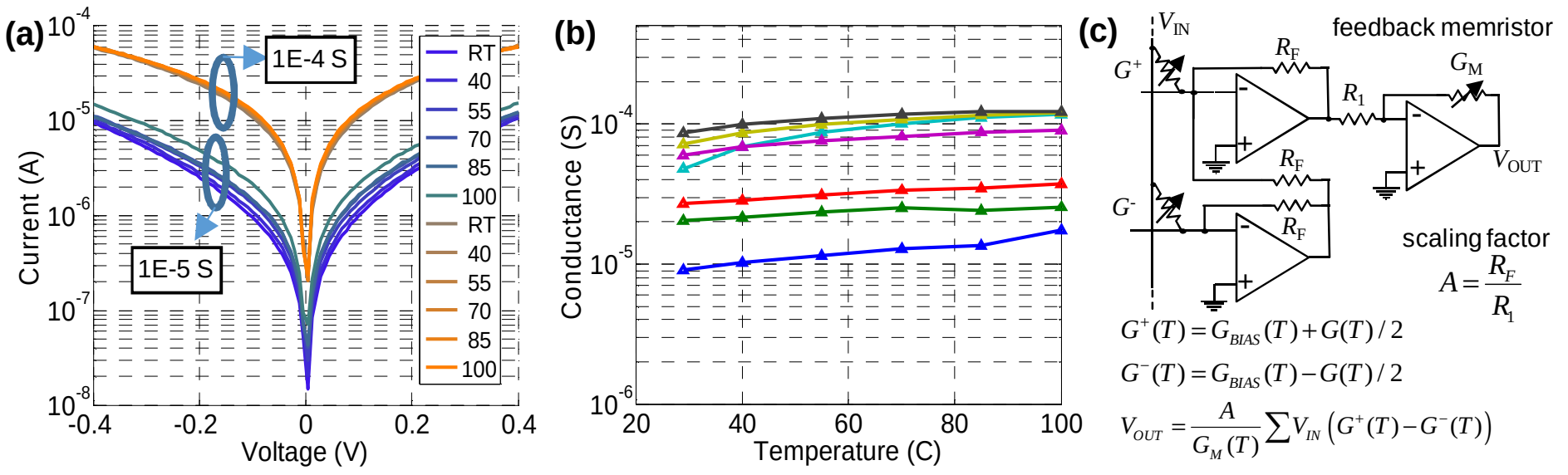

Fig. 13. Temperature sensitivity study: (a) The $I$-$V$ curves of a single memristor for several temperatures and (b) the extracted temperature dependence of its conductance. (c) The proposed vector-by-matrix multiplier circuit for wide-temperature-range operation. (For clarity, only one memristor pair from an array, and only one neuron are shown.) The drift in conductance over temperature is reduced by choosing larger values of $G_{BIAS}$ at which dependence is the smallest and by using memristor in output opamp feedback.